\documentclass{sig-alternate-10pt}
\usepackage{cite}
\usepackage{graphicx}
\usepackage{algorithm}
\usepackage{algorithmic}
\usepackage{color}
\usepackage{subfigure}
\usepackage{epsfig}
\usepackage{setspace}
\usepackage{float}

\title{On the Efficacy of Live DDoS Detection with Hadoop}

\author{Sufian Hameed, Usman Ali\\
       IT Security Labs,\\
       National University of Computer and Emerging Sciences (NUCES), Pakistan\\
       sufian.hameed@nu.edu.pk, k133023@nu.edu.pk
}

\begin{document}

\maketitle

\begin{abstract}

Distributed Denial of Service (DDoS) flooding attacks are one of the biggest challenges to the availability of online services today. These DDoS attacks overwhelm the victim with huge volume of traffic and render it incapable of performing normal communication or crashes it completely. If there are delays in detecting the flooding attacks, nothing much can be done except to manually disconnect the victim and fix the problem. With the rapid increase of DDoS volume and frequency, the current DDoS detection technologies are challenged to deal with huge attack volume in reasonable and affordable response time.

In this paper, we propose HADEC, a Hadoop based Live DDoS Detection framework to tackle efficient analysis of flooding attacks by harnessing MapReduce and HDFS. We implemented a counter-based DDoS detection algorithm for four major flooding attacks (TCP-SYN, HTTP GET, UDP and ICMP) in MapReduce, consisting of map and reduce functions. We deployed a testbed to evaluate the performance of HADEC framework for live DDoS detection. Based on the experiment we showed that HADEC is capable of processing and detecting DDoS attacks in affordable time.

\end{abstract}


\section{Introduction}
\label{sec:intro}

Distributed Denial of Service (DDoS) flooding attacks are one of the biggest concerns for security and network professionals. The first DDoS attack incident \cite{criscuolo2000distributed} was reported in 1999 by the Computer Incident Advisory Capability (CIAC). Since then, most of the DoS attacks are distributed in nature and they continue to grow in frequency, sophistication and bandwidth. The main aim of these attacks is to overload the victim's machine and make his services unavailable, leading to revenue losses.

Over the years DDoS has hit major companies and Internet infrastructures, incurring significant loss in revenues. Yahoo! experienced one of the first major DDoS flooding attacks that made their services offline for about 2 hours \cite{yahoo}. In October 2002, 9 of the 13 DNS root servers were shut down for an hour because of a DDoS flooding attack \cite{dns}. During the fourth quarter of 2010, a hacktivist group called \emph{Anonymous} orchestrated major DDoS flooding attacks and brought down the Mastercard, PostFinance, and Visa websites \cite{anonymous}. Most recently, online banking sites of 9 major U.S. banks (i.e., Bank of America, Citigroup, Wells Fargo, U.S. Bancorp, PNC, Capital One, Fifth Third Bank, BB\&T, and HSBC) have been continuously the targets of powerful DDoS flooding attack series [15]. The legacy of DDoS continue to grow in sophistication and volume with recent attacks breaking the barrier of 100 Gbps \cite{zargar2013survey}



The explosive increase in the volume of internet traffic and sophistication of DDoS attacks have posed serious challenges on how to analyze the DDoS attacks in a scalable and accurate manner. For example, two of the most popular open-source intrusion detection systems (IDS), Snort \cite{snort} and Bro \cite{bro}, maintain per-flow state to detect anomalies. The Internet traffic doubles every year and due to that monitoring large amount of traffic in real-time anomaly detection with conventional IDS has become a bottleneck.

In \cite{lee2011detecting}, Lee et al. has proposed a DDoS detection method based on Hadoop \cite{hadoop}. They have used a Hadoop based packet processor \cite{traceprocessing} and devised a MapReduce \cite{mapreduce} based detection algorithm against the HTTP GET flooding attack. They employ a counter-based DDoS detection algorithm in MapReduce that counts the total traffic volume or the number of web page requests for picking out attackers from the clients. For experiments, they used multiple Hadoop nodes (max. 10) in parallel to show the performance gains for DDoS detection. Unfortunately, their proposed framework, in its current form can only be used for offline batch processing of huge volume of traces. The problem to develop a real time defense system for live analysis still needs to be tackled.

In this paper, we propose HADEC, a Hadoop based Live DDoS Detection framework. HADEC is a novel destination based DDoS defense mechanism that leverages Hadoop to detect live DDoS flooding attacks in wired networked systems. HADEC comprise of two main components, a capturing server and a detection server. Live DDoS starts with the capturing of live network traffic handled by the capturing server. The capturing server then process the captured traffic to generate log file and transfer them to the detection server for further processing. The detection server manages a Hadoop cluster and on the receipt of the log file(s), it starts a MapReduce based DDoS detection jobs on the cluster nodes. The proposed framework implements counter-based algorithm to detect four major DDoS flooding attacks (TCP-SYN, UDP, ICMP and HTTP GET). These algorithms executes as a reducer job on the Hadoop detection cluster.

We also deploy a testbed for HADEC which consists of a capturing server, detection server and a cluster of ten physical machines, each connected via a Gigabit LAN. We evaluate HADEC framework for live DDoS detection by varying the attack volume and cluster nodes. HADEC is capable of analyzing 20 GB of log file, generated from 300 GBs of attack traffic, in approx. 8.35 mins on a cluster of 10 nodes. For small log files representing 1.8 Gbps the overall detection time is approx. 21 seconds.

The rest of the paper is organized as follows. \textsection\ref{sec:related} describes the state of the art. \textsection\ref{sec:framework} describes the HADEC framework design. In \textsection\ref{sec:eval} we discuss the testbed and demonstrate the performance of the proposed framework. Finally we conclude the paper in \textsection\ref{sec:conclusion}.

\section{Related Work}
\label{sec:related}

Since the inception of DDoS flooding attacks, several defense mechanisms have been proposed to date in the literature \cite{zargar2013survey}. This section highlights the defense mechanisms against two main DDoS flooding attacks, followed by a discussion on the application of Mapreduce/Hadoop to combat network anomalies, Botnet and DDoS related attacks.

The DDoS flooding attacks can be categorized into two types based on the protocol level that is targeted: network/transport-level attacks (UDP flood, ICMP flood, DNS flood, TCP SYN flood, etc.) and application-level attacks (HTTP GET/POST request). The defense mechanisms against network/transport-level DDoS flooding attacks roughly falls into four categories: \emph{source-based, destination-based, network-based, and hybrid (distributed)} and the defense mechanisms against application-level DDoS flooding attacks have two main categories: \emph{destination-based, and hybrid (distributed)}. Since the application traffic is not accessible at the layer 2 and layer 3, there is no network-based defense mechanism for the application-level DDoS. Following is the summary of features and limitations for the DDoS defense categories.

\begin{itemize}
  \item \textbf{Source-Based:} In source-based defense mechanism the detection and response are deployed at the source hosts in an attempt to mitigate the attack before it wastes lots of resources \cite{mirkovic2002attacking, mirkovic2003source}. Accuracy is a major concern in this approach as it is difficult to differentiate legitimate and DDoS attack traffic at the sources with low volume of the traffic. Further there is low motivation for deployment at the source ISP due to added cost for community service.
  \item \textbf{Destination-Based:} In this case the detection and response mechanisms are deployed at the destination hosts. Access to the aggregate traffic near the destination hosts makes the detection of DDoS attack easier and cheaper, with high accuracy, than other mechanisms \cite{wang2007defense, yaar2003pi, ranjan2009ddos}. On the downside destination based mechanisms cannot preempt a response to the attack before it reaches the victim and wastes resources on the paths to the victim.
  \item \textbf{Network-Based:} With network-based approach the detection and response are deployed at the intermediate networks (i.e., routers). The rational behind this approach is to filter the attack traffic at the intermediate networks and as close to source as possible \cite{park2001effectiveness, park2001}. Network-based DDoS defenses incur high storage and processing overhead at the routers and accurate attack detection is also difficult due to lack of sufficient aggregated traffic destined for the victims.
  \item \textbf{Hybrid (Distributed):} In hybrid approach there is coordination among different network components along the attack path and detection and response mechanisms are deployed at various locations.  Destination hosts and intermediate networks usually deploy detection mechanisms and response usually occurs at the sources and the upstream routers near the sources \cite{yang2008tva, yang2005limiting}. Hybrid approach is more robust against DDoS attacks, but due to distributed nature, it requires more resources at various levels (e.g., destination, source, and network) to tackle DDoS attacks. The complexity and overhead because of the coordination and communication among distributed components is also a limiting factor is smooth deployment of hybrid-based DDoS defenses.
\end{itemize}

Analysis of logs and network flows for anomaly detection has been a problem in the information security for decades. New big data technologies, such as Hadoop, has attracted the interest of the security community for its promised ability to analyze and correlate security-related heterogeneous data efficiently and at unprecedented scale and speeds \cite{cardenas2013big}. In the rest of the section, we review some recent techniques (other than \cite{lee2011detecting} , discussed in \textsection\ref{sec:intro}) where Hadoop based frameworks are used to build affordable infrastructures for security applications.

BotCloud \cite{francois2011botcloud} propose a scalable P2P detection mechanism based on MapReduce and combination of host and network approaches \cite{franccois2011bottrack}. First they generate large dataset of Netflow data \cite{claise2004cisco} on an individual operator. Next they applied a PageRank algorithm on the Netflow traces to differentiate the dependency of hosts connected in P2P fashion for the detection of botnets. They moved the pagerank algorithm to MapReduce and the pagerank algorithm executes on data nodes of Hadoop cluster for efficient execution.

Temporal and spatial traffic structures are essential for anomaly detectors to accurately drive the statistics from network traffic. Hadoop divides the data into multiple same size blocks, and distributes them in a cluster of data nodes to be processed independently. This could introduce a difficulty in analysis of network traffic where related packets may be spread across different block, thus dislocating traffic structures. Hashdoop \cite{fontugne2014hashdoop} resolve this potential weakness by using hash function to  divide traffic into blocks that preserve the spatial and temporal traffic structures. In this way, Hashdoop conserves all the advantages of the MapReduce model for accurate and efficient anomaly detection of network traffic.

\begin{figure}[h]
\centering
\includegraphics[width=0.4\textwidth]{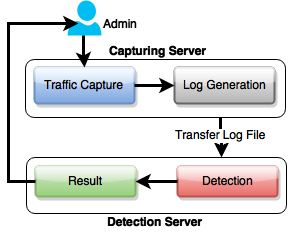}
\caption{Different Phases of HADEC}
\label{HADECPhases}
\end{figure}


\section{Hadoop DDoS Detection Framework}
\label{sec:framework}

The Hadoop Based Live DDoS Detection Framework (HADEC) comprise of four major phases (see fig.~\ref{HADECPhases}).
\begin{enumerate}
 \item Network traffic capturing and Log generation.
 \item Log transfer.
 \item DDoS detection.
 \item Result notification.
\end{enumerate}

Each of the above mentioned phases are implemented as separate components that communicate with each other to perform their assigned task. Traffic capturing and log generation are handled at the \emph{capturing server}, whereas DDoS detection and result notification is performed by the \emph{detection server}. Log transfer is handled through web services.  In the following subsections we have explained the functionalities for each of the phase/component in detail.

\begin{figure}[h]
\centering
\includegraphics[width=0.4\textwidth]{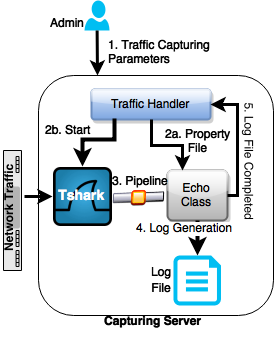}
\caption{Network Traffic Capturing and Log Generation Component}
\label{CapturePhase}
\end{figure}

\subsection{Traffic Capturing and Log Generation}
\label{sec:traffic_capturing}

live DDoS detection starts with the capturing of network traffic. HADEC provides a web interface through which the admin can tune the capturing server with desired parameters. These parameters are; file size, number of files to be captured before initializing the detection phase and the path to save the captured file. Once the admin is done with the configurations, the \emph{Traffic Handler} sends the property file to the \emph{Echo Class} (a java utility to generate logs) and start the capturing of live network traffic (see fig.~\ref{CapturePhase}).

HADEC use the Tshark library \cite{tshark} to capture live network traffic. Tshark is an open source library capable of capturing huge amount of traffic. Under default settings, Tshark library runs through command line, and outputs the result on console. To log the traffic for later use, we developed a java based utility (Echo Class) to create a pipeline with Tshark and read all the output packets from Tshark. We have also tuned Tshark to output only the relevant information required during detection phase. This includes information of timestamps, src IP, dst IP, packet protocol and brief packet header information. Following are the snippets for TCP (SYN), HTTP, UDP and ICMP  packets that are logged in the file.

\begin{verbatim}

TCP (SYN)
17956  45.406170  10.12.32.1 -> 10.12.32.101
TCP 119 [TCP Retransmission] 0 > 480 [SYN]
Seq=0 Win=10000 Len=43 MSS=1452 SACK_PERM=1
TSval=422940867 TSecr=0 WS=32

HTTP
46737 2641.808087 10.12.32.1 -> 10.12.32.101
HTTP 653 GET /posts/17076163/ivc/dddc?
_=1432840178190 HTTP/1.1

UDP
139875	138.04015 10.12.32.1 -> 10.12.32.101
UDP	50	Src port: 55348  Dst port: http

ICMP
229883	2658.8827  10.12.32.1 ->  10.12.32.1O1
ICMP	42	Echo (ping) request  id=0x0001,
seq=11157/38187, ttl=63 (reply in 229884)

\end{verbatim}

As discussed above, the \emph{Traffic Handler} sends the property file to the \emph{Echo Class} with the desired set of parameters (file size, file count for detection and storage path on the capturing server) set by the admin. Echo Class use these parameters to generate a log file, at the specified location, when it reads the required amount of data from Tshark. Once the log file is generated, the \emph{Echo Class} also notifies the \emph{Traffic Handler} (see fig.~\ref{CapturePhase}).

\subsection{Log Transfer Phase}

\begin{figure}[h]
\centering
\includegraphics[width=0.48\textwidth]{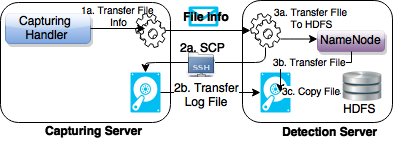}
\caption{Log Transfer Phase}
\label{TransferPhase}
\end{figure}

After the log file is generated, the \emph{Traffic Handler} in the capturing server will temporarily pause the traffic capturing operations of Tshark. The traffic handler will then notify the \emph{detection server} and also share the file information (file name, file path, server name, etc.) with it via a webservice. The detection server will initiate a Secure Copy or SCP protocol \cite{scp}(with pre-configured credentials) with the capturing server, and transfer the log file from the capturing server (using the already shared name/path information) into its local file system (see fig.~\ref{TransferPhase}).

Since the detection server mainly works as a NameNode i.e. the centerpiece of the Hadoop cluster and HDFS (Hadoop distributed file system), it has to transfer the log file(s) from local storage to HDFS. On successful transfer of log file into HDFS, the detection server sends a positive acknowledgement to the capturing server and both the servers delete that specific file from their local storage to maintain healthy storage capacity. On the receipt of successful log file transfer, the traffic handler will restart the Tshark for capturing network traffic. Before starting the DDoS detection process, the detection server will wait for the final acknowledgment from the capturing server. This acknowledgement validates that the desired number of files of a particular size (set via parameters by admin) has been transferred to HDFS before the execution of MapReduce based DDoS detection algorithm. There is no particular restriction on the minimum file count before the detection starts; it could be set to one.

\subsection{Detection Phase}

The Apache Hadoop consists of two core components i.e. HDFS (storage part) and MapReduce (processing part). Hadoop's central management node also known as NameNode splits the data into same size large blocks and distributes them amongst the cluster nodes (data nodes). Hadoop MapReduce transfers packaged code for nodes to process in parallel, the data each node is responsible to process.

\begin{figure}[h]
\centering
\includegraphics[width=0.48\textwidth]{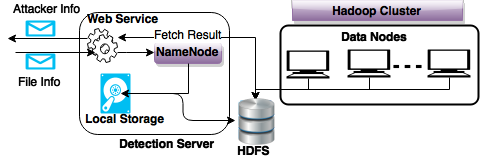}
\caption{DDoS Detection on Hadoop Cluster}
\label{DetectionPhase}
\end{figure}

In HADEC, the detection server mainly serves as the Hadoop's NameNode, which is the centerpiece of the Hadoop DDoS detection cluster. On successful transfer of log file(s), the detection server split the file into same size blocks and starts MapReduce DDoS detection jobs on cluster nodes (see fig.~\ref{DetectionPhase}). We have discussed MapReduce job analyzer and counter based DDoS detection algorithm in \textsection\ref{sec:mralgo}. Once the detection task is finished, the results are saved into HDFS.

\subsection{Result Notification}

\begin{figure*}[ht]
\centering
\includegraphics[width=0.9\textwidth]{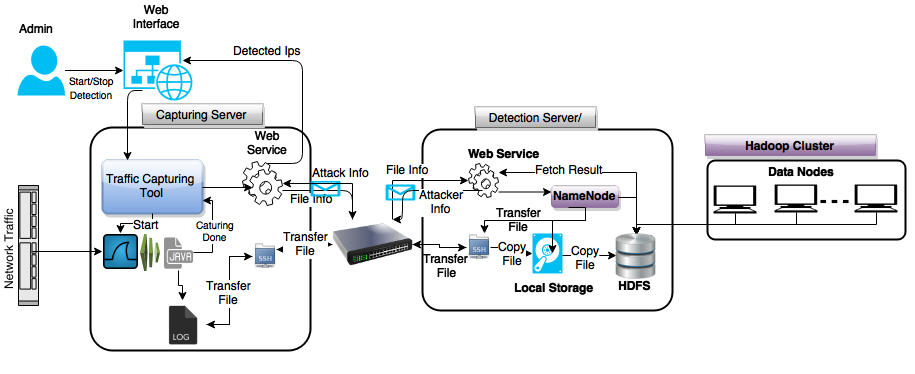}
\caption{HADEC: Hadoop Based DDoS Detection FrameWork}
\label{framework}
\end{figure*}

Once the execution of all the MapReduce tasks is finished, Hadoop will save the results in HDFS. The detection server will then parse the result file from HDFS and send the information about the attackers back to the administrator via the capturing server.  Once the results are notified both the input and output folders from HDFS will be deleted for better memory management by the detection server. Fig.~\ref{framework} presents a holistic illustration of HADEC framework.

\subsection{MapReduce Job and DDoS Detection}
\label{sec:mralgo}

A MapReduce program is composed of a Map task that performs filtering and sorting and a Reduce task that performs a summary operation. Here we have explained how HADEC has implemented detection of DDoS flooding attacks (UDP, HTTP GET, ICMP and TCP-SYN) as a MapReduce task on Hadoop cluster using \emph{counter-based algorithm}.

\subsubsection{HADEC Mapper job}

After starting MapReduce task, the first task is a mapper task which takes input from HDFS as a block. In our case the block will represent a file in text format and the input for each iteration of mapper function will be a single line from the file. Any single line in the file contains only brief information of a network packet captured through Tshark (see \textsection\ref{sec:traffic_capturing}). The term network packet used in the rest of this section represents a single line content of the file read as a mapper input.

Mapper job takes pair of data as input and returns a list of pairs (key, value). Mapper output type may differ from mapper's input type, in our case the input of mapper is pair of any number $i$ and the network packet. The output is a list of pair (key, value) with key as the src IP address and value as a network packet. Mapper job also use hashing for combining all the logs of data on the basis of src IP address, so that it becomes easier for reducer to analyze the attack traffic.

After all the mapper have finished their jobs, the data or worker nodes perform a shuffle step. During shuffling the nodes redistribute the data based on the output keys, such that all data belonging to one key is located on the same worker node (see fig. \ref{mapreduceTask}).

In HADEC, for analysis and detection of UDP flooding attack the mapper task filters out the packets having UDP information. In particular, the mapper function will search packets having QUIC / UDP information. QUIC stands for Quick UDP Internet connection. For the packet that contains the desired information, the mapper function generates an output in the form of pairs (key, value). The pseudocode for mapper function is as follows.

 \begin{verbatim}
%UDP detection mapper function
function Map is
  input: integer i, a network packet
begin function
  filter packet with QUIC/UDP type
  if packet does not contain information
  then
   ignore that packet
  else
   produce one output record (src ip, packet)
  end if
end function
\end{verbatim}

For ICMP, TCP-SYN and HTTP-GET based flooding attacks; the mapper function will search for SYN, ICMP and HTTP-GET packet type information respectively.

\subsubsection{HADEC Reducer job and Counter-Based Algorithm}

\begin{figure*}[ht]
\centering
\includegraphics[width=0.8\textwidth]{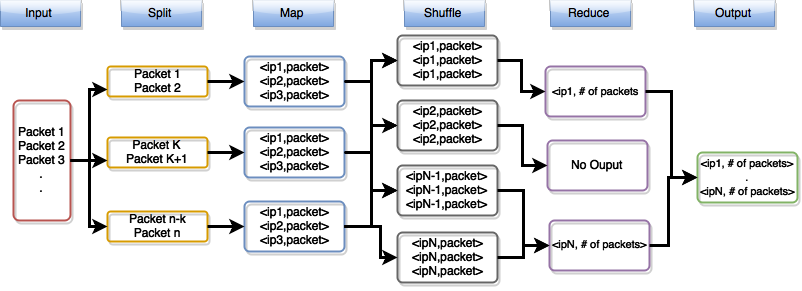}
\caption{Mapper and Reducer Operations}
\label{mapreduceTask}
\end{figure*}

Once the mapper tasks are completed, the reducer will start operating on the list of key/value pairs (i.e. IP/Packet pairs) produced by the mapper functions. The reducers are assigned a group with unique key, it means that all the packets with unique key (unique src IP in our case) will be assigned to one reducer. We can configure Hadoop to run reducer jobs on varying number of data nodes. For efficiency and performance it is very important to identify the correct number of reducers required for finalizing the analysis job. HADEC run counter-based algorithm to detect DDoS flooding attacks on reducer nodes. The reducer function takes input in key/value pair (srp IP, Packet of Type X) and produces a single key/value pair (src IP, No. of packets of type X) output after counting the number instance (see fig. \ref{mapreduceTask}).

\emph{Counter based algorithm} is the simplest, yet very effective algorithm to analyze the DDoS flooding attacks by monitoring the traffic volumes for src IP addresses. The algorithm counts all the incoming packets, of a particular type (UDP, ICMP, HTTP ...etc), associated with a unique IP address in a unit time. If the traffic volume or count for src IP exceeds the pre-defined threshold, that particular IP will be declared as an attacker. The pseudocode for reducer function using counter-based algorithm for UDP attack is as follows.

\begin{verbatim}
/* %Reducer function for UDP attack detection */
function Reduce is
  input: <source ip, UDP Packets>
  begin function
    count :=count # of packets for src IP
    if(count is greater than THRESHOLD)
      begin if
      /* This ip declares to be the Attacker ip */
        produce one ouput <Src IP, # of Packets>
      end if
    else
      Ignore (do nothing)
end function
\end{verbatim}

\section{TestBed and Evaluations}
\label{sec:eval}

In this section we have discussed the testbed deployment of HADEC and how we have evaluated the performance of the proposed framework with different experiment.

\subsection{HADEC TestBed}

HADEC perform two main tasks, (a) capturing and transfer of network traffic and (b) detection of DDoS flooding attacks. For capturing the traffic we use a single node capturing server to capture, process and send the network traffic to detection server. For DDoS detection, we deploy a single node detection server (also acts as NameNode of Hadoop cluster) and a Hadoop detection cluster consisting of ten nodes. Each node in our testbed (one capturing server, one detection server and ten Hadoop data nodes) consists of 2.60 GHz Intel core i5 CPU, 8 GB RAM, 500 GB HDD and 1 Gbps Ethernet card. All the nodes in HADEC used Ubuntu 14.04 and are connected over a Gigabit LAN. We have used Hadoop version 2.6.0 for our cluster and YARN \cite{yarn} to handle all the JobTracker and TaskTracker functionality.

\begin{figure}[h]
\centering
\includegraphics[width=0.48\textwidth]{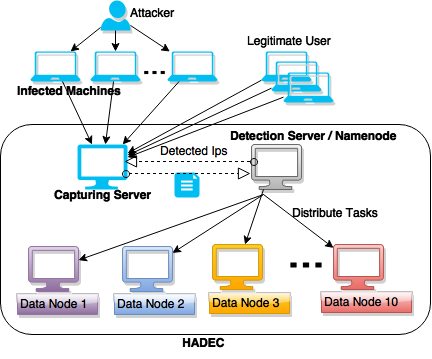}
\caption{HADEC Testbed }
\label{HADEC_TestBed}
\end{figure}

There are several attack generation tools that are available online, such as LOIC \cite{loic}, Scapy \cite{scapy}, Mausezahn \cite{mausezahn}, Iperf \cite{Iperf}, etc. For our testbed evaluations we have mainly used Mausezahn, because of its ability to generate huge amount of traffic with random IPs to emulate different number of attackers. We deployed three dedicated attacker nodes along with couple of legitimate users to flood the victim machine (capturing server) with a traffic volume of uptil 913Mbps (practically highest possible for a Gigabit LAN). HADEC testbed is shown in fig. \ref{HADEC_TestBed}. For evaluations we have only focused on UDP flooding attack due to its tendency to reach high volume from limited number of hosts. We would also like to add that for all the evaluations we have used only a single reducer, different variations were tried but but there was no performance gains.

\subsection{Performance Evaluation}

The overall performance of HADEC depends on the time taken for capturing the log file(s) at the capturing server, transferring the file to the detection server and execution of counter-based DDoS detection algorithm on the Hadoop cluster. For our evaluations, we varied different parameters like; log file size, Hadoop cluster size, Hadoop splits or block size and threshold for counter-based algorithm, and measured their impact on the performance of HADEC.

\subsubsection{Traffic Capturing and File Transfer}

The capturing server work on two major tasks simultaneously. First, it captures huge amount of network traffic (913 Mbps in our testbed) and transform it into log file(s). Second it transfers the log file(s) to detection server for further processing. This simultaneous execution of capture and transfer operations are important for live analysis of DDoS flooding attack, but on the hand both the operations compete for resources.

\begin{figure}[h]
\centering
\includegraphics[width=0.46\textwidth]{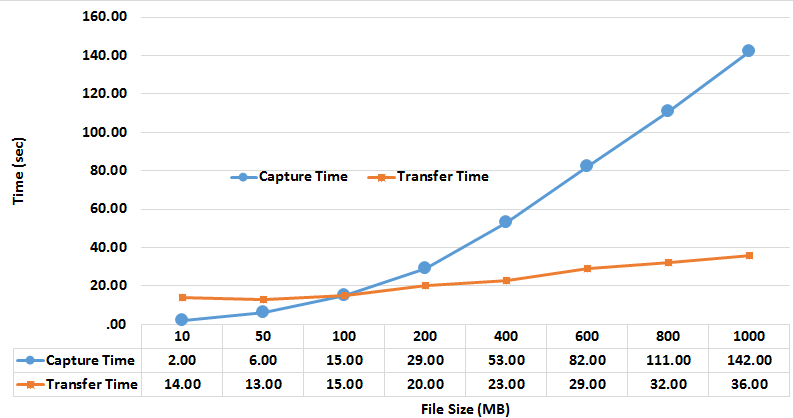}
\caption{Capture and transfer time of a log file.}
\label{captureTransfer}
\end{figure}

Fig. \ref{captureTransfer} shows the capturing and transfer time taken by the capturing server for log files of different sizes. The capturing time is almost linear to the increase in file size. It takes approx 2 seconds to log a file of 10 MB and extends to 142 seconds for 1 GB file. File transfer takes 14 seconds to transfer 10 MB file and approx. 35 seconds for 1GB file. This shows a clear improvement in throughput with the increase in file size. Here it is also interesting to note that the transfer operation has to compete for bandwidth and during peak time more than 90\% of the bandwidth is being consumed by the attack traffic.

\subsubsection{Number of Attackers and Attack Volume}

Table. \ref{volume} presents the relationship between the size of log file with the total number of attackers and aggregate traffic volume. HADEC use counter-based algorithm to detect attackers. This means that during the DDoS flooding attack, any particular attacker has to cross certain volume threshold to be detected. According to the table. \ref{volume}, the capturing server has to analyze approx. 0.24 GBs of network traffic to generate a log file of 10 MB and it could represent 100 plus attacker that cross the flooding frequency threshold of 500-1000 packet. By increasing the log file size, the capability to capture accurate information related to attackers also increases. There is a trade-off between the log file size and overall detection rate, therefore, the admin will have to adjust the framework parameters that will best fit in different attack scenarios.

\begin{table}
\centering
\caption{Relationship of Log File Size with No. of Attackers and Traffic Volume}
\vspace{4mm}
\begin{tabular}{|l|l|l|}
  \hline  File Size (MB) & No. Of Attackers & Traffic Vol. \\
  \hline  10 & 100 & 0.22 GB\\
  \hline  50 & 500 & 0.67 GB\\
  \hline  100 & 1500 & 1.67 GB\\
  \hline  200 & 2000 & 3.23 GB\\
  \hline  400 & 4000 & 5.91 GB\\
  \hline  600 & 6000 & 9.14 GB\\
  \hline  800 & 8000 & 12.37 GB\\
  \hline  1000 & 10,000 & 15.83 GB\\
  \hline
\end{tabular}

\label{volume}
\end{table}

\subsubsection{DDoS Detection on Hadoop Cluster}


We evaluate the performance of DDoS detection phase on the basis of different size of the log files, different data block size for MapReduce tasks, different threshold value for counter-based detection algorithm. For our evaluations we used one fix 80-20 attack volume (80\% attack traffic and 20\% legitimate traffic). We have used these setting to emulate flooding behavior where attack traffic surpass the legitimate one.

\begin{figure}[H]
\centering
\includegraphics[width=0.48\textwidth]{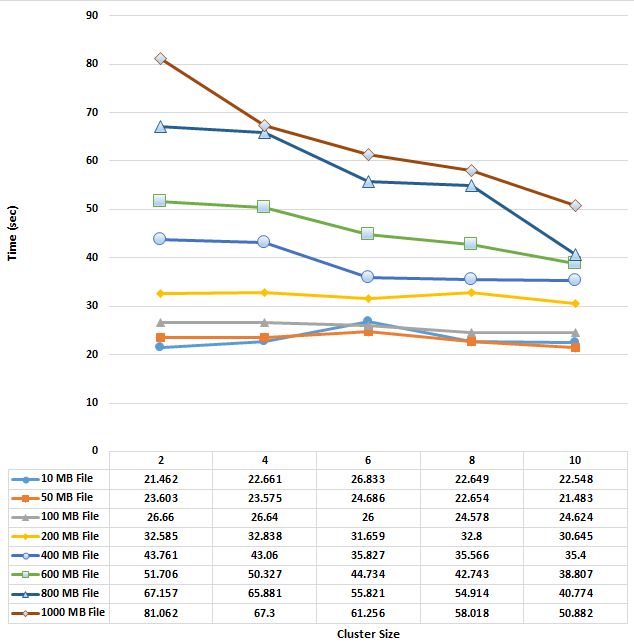}
\caption{Detection time at Hadoop cluster with 500 threshold}
\label{DTimeThresh500V80-20}
\end{figure}

Fig. \ref{DTimeThresh500V80-20} shows the detection time on Hadoop cluster. In this experiment we used a fix threshold of 500 and data block of 128 MB. Detection is performed based on different file size and varying number of cluster nodes. With the increase in file size the number of attack traffic also increases, which affects the mapper and reducer operation frequency and time. In short with the increase in file size the detection time increase and it will also increase the detection rate or the number of attackers IPs, which is a plus point. Increase in cluster size hardly effects the detection time for files less 400 MB in size, on the contrary in some cases it might increase a little due to added management cost. Hadoop enables parallelism by splitting the files into different blocks of specified size. Files smaller than the Hadoop block size are not split over multiple nodes for execution. Therefore, the overall detection time remains the same over different cluster node.

Starting from the file size of 400 MB, the detection time improves with the increase of cluster size. For bigger files like 800 MB and 1000 MB, Hadoop work more efficiently. We can see that the detection time reduces around 27 to 30 for 800 and 1000 MB files respectively, when the cluster size is increased from 2 to 10 nodes. This is because with 1000MB file there are 9 blocks and with the increase in cluster size, Hadoop will assign the task to different nodes in parallel.

\begin{figure}[H]
\centering
\includegraphics[width=0.41\textwidth]{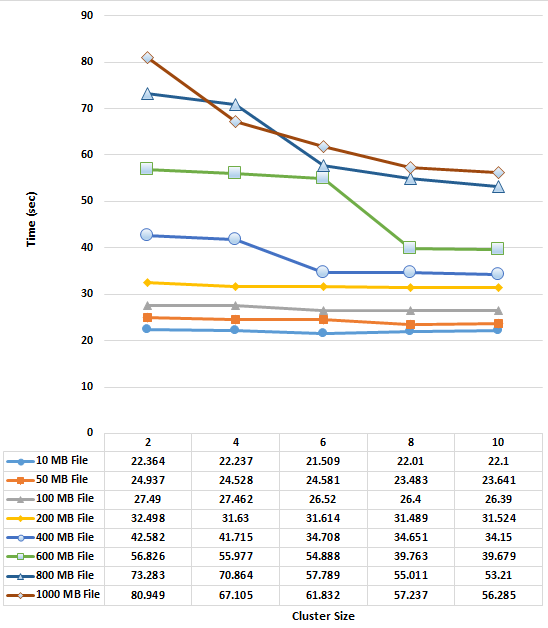}
\caption{Detection time at Hadoop cluster with 1000 threshold}
\label{DTimeThresh1000V80-20}
\end{figure}

Fig. \ref{DTimeThresh1000V80-20} shows the detection time on Hadoop cluster with a threshold value of 1000. In this experiment we only change the threshold value and all the remaining settings are similar to the fig. \ref{DTimeThresh500V80-20}. With the increase in threshold value the total number of inputs for reducers also increases and this will increase the reducer time. This is the reason why \emph{majority} of results in shown in fig. \ref{DTimeThresh1000V80-20} has couple of seconds higher detection time as compared to the results in fig. \ref{DTimeThresh500V80-20}.

\begin{figure}[H]
\centering
\includegraphics[width=0.41\textwidth]{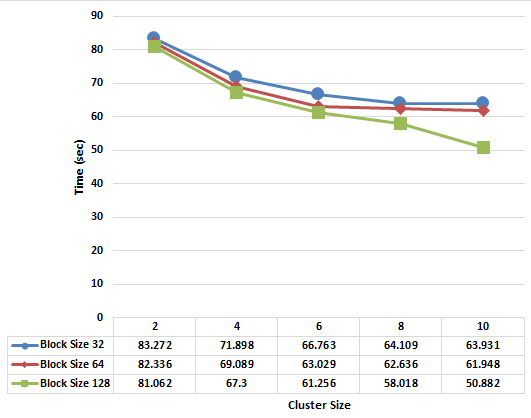}
\caption{Detection time with different block sizes on 1 GB file }
\label{1000GBllBlockSizes}
\end{figure}

Fig. \ref{1000GBllBlockSizes} shows the effect of varying block sizes on the detection time for 1 GB file. In this experiment we use fix threshold of 500 and use three different blocks of size 32, 64 and 128 MB. For 1 GB file the block size of 128 MB gives the maximum performance gains in terms of detection time with the increase in cluster nodes. With smaller block size there are more splits, resulting in multiple tasks being schedule on a mapper and adds management overhead.

\begin{figure}[h]
\centering
\includegraphics[width=0.46\textwidth]{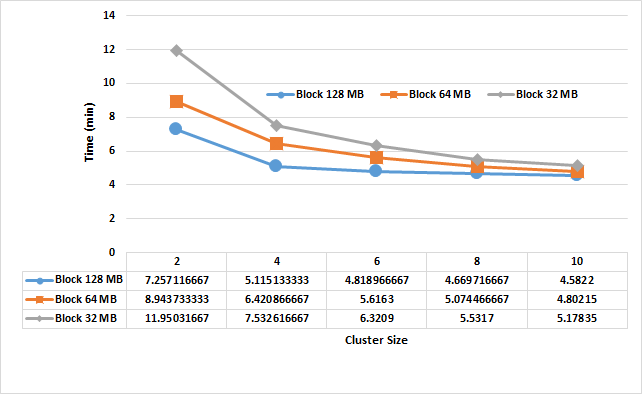}
\caption{Effect of block size on 10 GB file}
\label{10GB}
\end{figure}

\begin{figure}[H]
\centering
\includegraphics[width=0.46\textwidth]{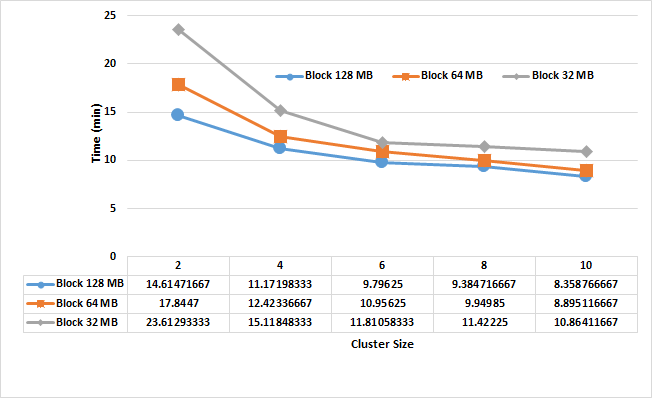}
\caption{Effect of block size on 20 GB file}
\label{20GB}
\end{figure}

The effect of cluster size is prominent on large files. This is because with large files, Hadoop can effectively split the files in multiple blocks and distributed on the available cluster nodes. Fig. \ref{10GB} and \ref{20GB} shows the effect of different block size and cluster node on the detection time, with a fix threshold of 500 and 80-20 attack volume. 128 MB block size gives the most efficient results; this is because when the number of blocks increases the resource manager in Hadoop needs to manage each of the blocks and its result. Thus, it will take more time to manage each task. For larger block size there is only one map task to process the whole large block. On a 10 GB file with a block size of 128 MB, Hadoop finished the detection task in approx. 7.5 mins with a cluster size of 2 nodes. The detection time goes down to approx 4.5 mins when the cluster size is increased to 10 nodes. For 20 GB file with a block size of 128 MB, the time to finish the detection task is 14.6 mins and 8.3 mins on a cluster of 2 and 10 nodes respectively. If we approximate the numbers in table. \ref{volume}, HADEC can effectively resolve 100K attackers for an aggregate traffic volume of 159 GBs with 10 GB of log file in just 4.5 mins. These numbers will be doubled for 20 GB.

\subsubsection{Overall Framework Performance}

\begin{figure}[H]
\centering
\includegraphics[width=0.4\textwidth]{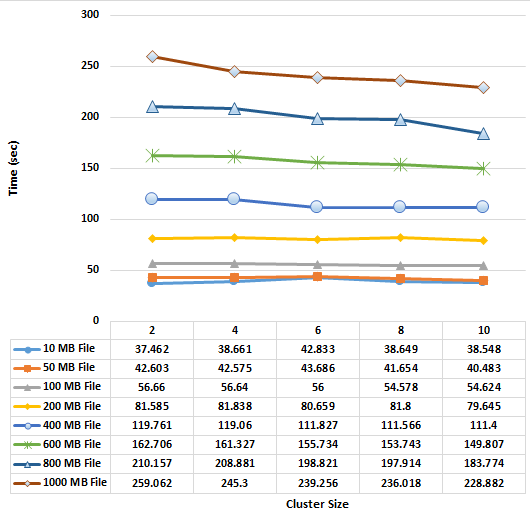}
\caption{Total time to detection DDoS Attack in HADEC with 500 threshold}
\label{TotalTimeThresh500V80-20}
\end{figure}
\begin{figure}[H]
\centering
\includegraphics[width=0.4\textwidth]{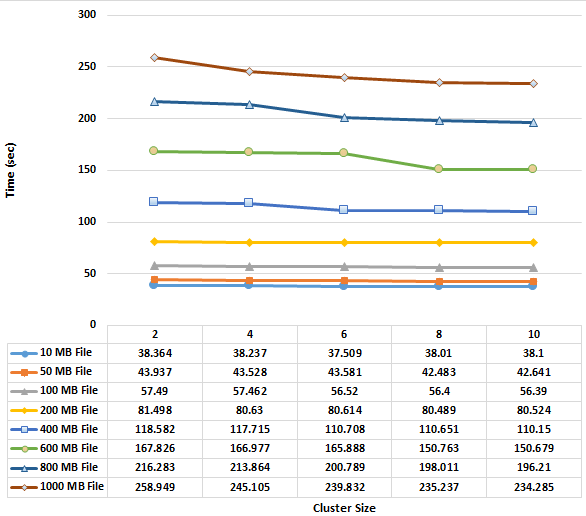}
\caption{Total time to detection DDoS Attack in HADEC with 1000 threshold}
\label{TotalTimeThresh1000V80-20}
\end{figure}

Fig. \ref{TotalTimeThresh500V80-20} and \ref{TotalTimeThresh1000V80-20} shows the overall performance of our proposed framework to detect the DDoS attacks. These numbers present the total time required for capturing, processing, transferring and detection with different file sizes. For the experiments in fig. \ref{TotalTimeThresh500V80-20} we have used 80-20 attack volume, 128 MB block size and 500 threshold. For the experiments in fig. \ref{TotalTimeThresh1000V80-20}, we have only changed the threshold to 1000. In fig. \ref{TotalTimeThresh500V80-20}, we can observe that with the increase in the file size, the overall overhead of capturing and transferring phase increase. A 10 MB file takes approx. 16 seconds (42\%) in capturing/tranferring phase and 21 seconds in detection phase.  The best case of 1 GB file (10 node cluster) takes 178 seconds (77\%) in capturing/tranferring phase and just 50 seconds in detection phase. On the whole, it takes somewhere between 4.3 mins to 3.82 mins to analyze 1 GB of log file that can resolve 10K attackers and generated from an aggregate attack volume of 15.83 GBs.

\subsection{Discussion}

Based on the results presented in this section, we can conclude that HADEC is capable of analyzing huge volume of DDoS flooding attacks in scalable manner. Several GBs (1 GB file generated from 15.83 GBs of live traffic) of attack traffic can be analyzed in less than 5 mins.  By using small size for log file the overall detection time can be reduced to couple of seconds (30-40 seconds). But small log files also have an inherent limitation to the number of attacker's they can track. There is no single recommended setting; the admin will have to tweak the framework configuration that best match their requirement.

We also noticed that with smaller files, Hadoop does not provide parallelism. This means that if any admin configures HADEC to work on small files of under 800 MB, there will be no point in setting up multiple node cluster. A single or two node cluster of Hadoop will do the job within few minutes (2-3) with the hardware settings we used in our testbed. In our evaluations of HADEC, capturing and transferring phase showed the performance overhead and majority of the framework time was spent in these phases. This problem could be easily resolved by using reasonable to high-end server optimized for traffic operations, instead of mid-level core i5 desktop that are used in our testbed.

\section{Conclusions}
\label{sec:conclusion}

In this paper, we present HADEC, a scalable Hadoop based Live DDoS Detection framework that is capable of analyzing DDoS attacks in affordable time. HADEC captures live network traffic, process it to log relevant information in brief form and use MapReduce and HDFS to run detection algorithm for DDoS flooding attacks. HADEC solve the scalability, memory inefficiency and process complexity issues of conventional solution by utilizing parallel data processing promised by Hadoop. The evaluation results showed that HADEC would less than 5 mins to process (from capturing to detecting) 1 GB of log file, generated from approx. 15.83 GBs of live network traffic. With small log file the overall detection time can be further reduced to couple seconds.

We have observed that capturing of live network traffic incur the real performance overhead for HADEC. In worse case the capturing phase consumes 77\% of the overall detection time. As a future work, HADEC framework may allow potential optimizations to improve the capturing efficiency.

\bibliographystyle{plain}

\bibliography{DDoS}

\begin{thebibliography}{10}

\bibitem{hadoop}
Hadoop.
\newblock https://hadoop.apache.org/.

\bibitem{yarn}
Hadoop yarn.
\newblock http://hortonworks.com/hadoop/yarn/.

\bibitem{Iperf}
Iperf: network performance measurement tool.
\newblock https://iperf.fr/.

\bibitem{loic}
Loic: A network stress testing application.
\newblock http://sourceforge.net/projects/loic/.

\bibitem{mapreduce}
Mapreduce.
\newblock http://wiki.apache.org/hadoop/MapReduce.

\bibitem{mausezahn}
Mausezahn.
\newblock http://www.perihel.at/sec/mz/.

\bibitem{anonymous}
Operation payback cripples mastercard site in revenge for wikileaks ban, dec.
  8, 2010, [online].
\newblock http://www.guardian.co.uk.

\bibitem{dns}
Powerful attack cripples internet, oct. 23, 2002, [online].
\newblock http://www.greenspun.com/.

\bibitem{scapy}
Scapy.
\newblock http://www.secdev.org/projects/scapy/.

\bibitem{scp}
Secure copy.
\newblock linux.die.net/man/1/scp.

\bibitem{tshark}
Tshark: Network analyzer.
\newblock www.wireshark.org/docs/man-pages/tshark.html.

\bibitem{yahoo}
Yahoo on trail of site hackers, wired.com, feb. 8, 2000, [online].
\newblock http://www.wired.com/.

\bibitem{cardenas2013big}
Alvaro~A Cardenas, Pratyusa~K Manadhata, and Sreeranga~P Rajan.
\newblock Big data analytics for security.
\newblock {\em IEEE Security \& Privacy}, (6):74--76, 2013.

\bibitem{claise2004cisco}
Benoit Claise.
\newblock Cisco systems netflow services export version 9, rfc 3954
  (informational).
\newblock 2004.

\bibitem{criscuolo2000distributed}
Paul~J Criscuolo.
\newblock Distributed denial of service: Trin00, tribe flood network, tribe
  flood network 2000, and stacheldraht ciac-2319.
\newblock Technical report, DTIC Document, 2000.

\bibitem{fontugne2014hashdoop}
Romain Fontugne, Johan Mazel, and Kensuke Fukuda.
\newblock Hashdoop: A mapreduce framework for network anomaly detection.
\newblock In {\em Computer Communications Workshops (INFOCOM WKSHPS), 2014 IEEE
  Conference on}, pages 494--499. IEEE, 2014.

\bibitem{francois2011botcloud}
Jerome Francois, Shaonan Wang, Walter Bronzi, R~State, and Thomas Engel.
\newblock Botcloud: Detecting botnets using mapreduce.
\newblock In {\em Information Forensics and Security (WIFS), 2011 IEEE
  International Workshop on}, pages 1--6. IEEE, 2011.

\bibitem{franccois2011bottrack}
Jerome Francois, Shaonan Wang, Radu State, and Thomas Engel.
\newblock Bottrack: Tracking botnets using netflow and pagerank.
\newblock In Jordi Domingo-Pascual, Pietro Manzoni, Sergio Palazzo, Ana Pont,
  and Caterina Scoglio, editors, {\em NETWORKING 2011}, volume 6640 of {\em
  Lecture Notes in Computer Science}, pages 1--14. Springer Berlin Heidelberg,
  2011.

\bibitem{traceprocessing}
Yeonhee Lee, Wonchul Kang, and Youngseok Lee.
\newblock A hadoop-based packet trace processing tool.
\newblock In Jordi Domingo-Pascual, Yuval Shavitt, and Steve Uhlig, editors,
  {\em Traffic Monitoring and Analysis}, volume 6613 of {\em Lecture Notes in
  Computer Science}, pages 51--63. Springer Berlin Heidelberg, 2011.

\bibitem{lee2011detecting}
Yeonhee Lee and Youngseok Lee.
\newblock Detecting ddos attacks with hadoop.
\newblock In {\em Proceedings of The ACM CoNEXT Student Workshop}, page~7. ACM,
  2011.

\bibitem{mirkovic2002attacking}
Jelena Mirkovic, Gregory Prier, and Peter Reiher.
\newblock Attacking ddos at the source.
\newblock In {\em Proceedings of 10th IEEE International Conference on Network
  Protocols (ICNP)}, pages 312--321. IEEE, 2002.

\bibitem{mirkovic2003source}
Jelena Mirkovic, Gregory Prier, and Peter Reiher.
\newblock Source-end ddos defense.
\newblock In {\em Second IEEE International Symposium on Network Computing and
  Applications (NCA 2003)}, pages 171--178. IEEE, 2003.

\bibitem{park2001effectiveness}
Kihong Park and Heejo Lee.
\newblock On the effectiveness of probabilistic packet marking for ip traceback
  under denial of service attack.
\newblock In {\em Proceedings of 12th IEEE INFOCOM}, volume~1, pages 338--347.
  IEEE, 2001.

\bibitem{park2001}
Kihong Park and Heejo Lee.
\newblock On the effectiveness of route-based packet filtering for distributed
  dos attack prevention in power-law internets.
\newblock In {\em ACM SIGCOMM Computer Communication Review}, volume~31, pages
  15--26. ACM, 2001.

\bibitem{bro}
Vern Paxson.
\newblock Bro: A system for detecting network intruders in real-time.
\newblock {\em Comput. Netw.}, 31(23-24):2435--2463, December 1999.

\bibitem{ranjan2009ddos}
Supranamaya Ranjan, Ram Swaminathan, Mustafa Uysal, Antonio Nucci, and Edward
  Knightly.
\newblock Ddos-shield: Ddos-resilient scheduling to counter application layer
  attacks.
\newblock {\em IEEE/ACM Transactions on Networking (TON)}, 17(1):26--39, 2009.

\bibitem{snort}
Martin Roesch.
\newblock Snort - lightweight intrusion detection for networks.
\newblock In {\em Proceedings of the 13th USENIX Conference on System
  Administration}, LISA '99, pages 229--238, Berkeley, CA, USA, 1999. USENIX
  Association.

\bibitem{wang2007defense}
Haining Wang, Cheng Jin, and Kang~G Shin.
\newblock Defense against spoofed ip traffic using hop-count filtering.
\newblock {\em IEEE/ACM Transactions on Networking (ToN)}, 15(1):40--53, 2007.

\bibitem{yaar2003pi}
Abraham Yaar, Adrian Perrig, and Dawn Song.
\newblock Pi: A path identification mechanism to defend against ddos attacks.
\newblock In {\em Proceedings of IEEE Symposium on Security and Privacy}, pages
  93--107. IEEE, 2003.

\bibitem{yang2005limiting}
Xiaowei Yang, David Wetherall, and Thomas Anderson.
\newblock A dos-limiting network architecture.
\newblock In {\em ACM SIGCOMM Computer Communication Review}, volume~35, pages
  241--252. ACM, 2005.

\bibitem{yang2008tva}
Xiaowei Yang, David Wetherall, and Thomas Anderson.
\newblock Tva: a dos-limiting network architecture.
\newblock {\em IEEE/ACM Transactions on Networking}, 16(6):1267--1280, 2008.

\bibitem{zargar2013survey}
Saman~Taghavi Zargar, James Joshi, and David Tipper.
\newblock A survey of defense mechanisms against distributed denial of service
  (ddos) flooding attacks.
\newblock {\em Communications Surveys \& Tutorials, IEEE}, 15(4):2046--2069,
  2013.

\end{thebibliography}


\end{document}